# Phase Separation and Chemical Inhomogeneity in the Iron Chalcogenide Superconductor $Fe_{1+y}Te_xSe_{1-x}$

Hefei Hu, [1,3] J. M. Zuo, [2,3] J. S. Wen, [4] Z. J. Xu, [4] Z. W. Lin, [4] Q. Li, [4] Genda Gu, [4] W. K. Park, [1,3] and L. H. Greene[1,3]

[1]*Department of Physics, University of Illinois at Urbana-Champaign, Urbana, Illinois 61801, USA*

[2]*Department of Materials Science and Engineering, University of Illinois at Urbana-Champaign, Urbana, Illinois 61801, USA*

[3]*Frederick Seitz Materials Research Laboratory, University of Illinois at Urbana-Champaign, Urbana, Illinois 61801, USA*

[4]*Condensed Matter Physics and Materials Science Department, Brookhaven National Laboratory, Upton, New York 11973, USA*

Abstract: We report investigation on $Fe_{1+y}Te_xSe_{1-x}$ single crystals by using scanning transmission electron microscopy (STEM) and electron energy loss spectroscopy (EELS). Both non-superconducting samples with excess iron and superconducting samples demonstrate nanoscale phase separation and chemical inhomogeneity of Te/Se content, which we attribute to a miscibility gap. The line scan EELS technique indicates ~20% or less fluctuation of Te concentration from the nominal compositions.

Since the superconductivity in $LaFeAsO_{1-x}F_x$ was discovered in 2008[1], superconductivity in other iron-based compounds has attracted a lot of attention. Hsu et al. reported superconductivity at 8 K in PbO-type $FeSe$[2]. It has the simplest crystal structure, comprised of FeSe layers only, which is ideal for investigating the underlying mechanism in superconductivity. Furthermore, Te substitution to the Se sites of FeSe, i.e., $FeTe_xSe_{1-x}$, enhances the $T_c$ reaching ~15K at the optimized doping of 50% Te in bulk materials[3]. Further increase of the Te content lowers $T_c$[3], and the end member of this series, FeTe, shows long-range antiferromagnetic ordering with suppression of superconductivity[4]. The superconducting temperature transition width $\Delta T_c$ in $FeTe_xSe_{1-x}$ is often greater than 1 K as reported[3,5]; the origin of this width is not known. It has also been reported that the superconductivity in $FeTe_xSe_{1-x}$ is sensitive to the excess Fe which suppresses superconductivity[6,7]. For better understanding these materials, study on microstructure becomes essential. Scanning transmission electron microscopy (STEM) is an ideal tool for this purpose. In STEM incoherently scattered intensities, collected by the annular dark field (ADF) detector with a large inner cutoff angle, are proportional to atomic numbers, i.e., heavier atoms show brighter contrast than lighter atoms in ADF images. Electron energy loss spectroscopy (EELS), providing chemical information, can be recorded simultaneously in STEM mode. The power of STEM-EELS techniques has been demonstrated in FeAs-based superconductors[8]. In this letter, we present an investigation on the single crystals $Fe_{1+y}Te_xSe_{1-x}$ of four different nominal compositions, $Fe_{1.07}Te_{0.75}Se_{0.25}$ (non-superconductor), $Fe_{1.08}Te_{0.55}Se_{0.45}$ (non-superconductor), $FeTe_{0.7}Se_{0.3}$ ($T_c \sim 10$ K) and $FeTe_{0.55}Se_{0.45}$ ($T_c \sim 15$ K), by using STEM-EELS techniques. We show evidence of phase separation and chemical inhomogeneity in Te/Se content, which we attribute to a miscibility gap, in all of the samples, and use EELS to quantify the fluctuation of Te content.

Single crystals of $Fe_{1+y}Te_xSe_{1-x}$ were grown by a unidirectional solidification method with slow cool down[9]. The crystals have the morphology of thin flakes with [001] as the normal direction. To obtain TEM a specimen with uniform thickness in a relative large area (~ 100 nm), a method based on water dissolved tape is introduced. Due to the layer structure of iron-based superconductor, a thin piece of material can be peeled off from the single crystal by using adhesive tape (3M Mask Plus II—Water Soluble Wave Solder Tape). This thin piece of material is further exfoliated with tape to achieve thickness favorable for STEM-EELS measurement. A TEM copper grid with lacey carbon film is put on the adhesive side of tape from the final exfoliation, which is then immersed in deionized water at 70 °C. The tape on the TEM grid is dissolved and the specimens stay on lacey carbon film. Following further cleaning in water a few more times, the TEM grid is cleaned in acetone and isopropyl alcohol to remove residue on the specimens.

The microscope used was JEOL 2200FS (JEOL, USA) with a CEOS spherical aberration corrector (CEOS GmbH), operated at 200 keV with an electron probe that can be focused to 1 Å on the specimen[10]. Figure 1 shows an ADF image of the superconducting sample $FeTe_{0.7}Se_{0.3}$, showing the atomic structure projected on c axis. Te/Se atomic columns marked in blue, show brighter contrast than Fe atomic columns marked in red, since the atomic number of Te/Se is larger than that of Fe. Figure 2(a) illustrates an ADF image at atomic resolution at a medium magnification (×2M) of $FeTe_{0.7}Se_{0.3}$. It demonstrates clear contrast of interconnected, fractal morphologies. The region marked by a square in Figure 2(a) is magnified in Figure 2(b) with bright dots representing Te/Se atomic columns, showing coherent interfaces between bright and dark regions. Small lattice distortion is due to sample drifting. This contrast cannot be attributed to sample thickness variation because of the large uniform thickness achieved by the tape method we used for TEM sample preparation. Therefore, the appropriate interpretation on the contrast in ADF images shown in Fig. 2 would be that the brighter regions contain higher Te concentration than darker regions, since ADF image intensity is proportional to atomic numbers. Also, we noticed that the contrast remained similar when the sample was tilted by as much as 20°, and the thickness of samples we investigated ranges from 20 to 40 nm as estimated by the EELS log-ratio method[11], indicating that the Te rich or poor regions do not persist in the entire sample thickness, otherwise tilting angle of 20° would cause some of these regions to overlap in projection. All of other samples demonstrate similar topologies in recorded ADF images.

STEM-EELS experiments were performed for quantitative analysis on Te content. The condenser aperture was chosen to form a probe with convergence angle of 26 mrad, and the collection angle limited by entrance aperture was 32 mrad. Figure 3(a) shows an ADF image at atomic resolution recorded for the EELS measurement. The dashed line across the inhomogeneous regions on the image indicates the path along which the EELS spectra were recorded at an interval of 1.2 nm. The black solid spectrum in Figure 3(b) demonstrates a typical energy loss spectrum. The broad peak with the maximum intensity at around 625 eV corresponds to Te-$M_{4,5}$ edge, and the sharp peaks at 708 eV and 720 eV are Fe-$L_{2,3}$ edge. The red dotted spectrum in Fig. 3(b) shows the background subtracted spectrum for Te-$M_{4,5}$ edge, taking the signals from 532 eV to 570 eV as the background which was fitted by a power law. The background subtracted intensities of Te-$M_{4,5}$ edge were then integrated for the individual spectrum along the scanning path. The histogram in Fig. 3(c) shows the integrated Te-$M_{4,5}$ intensity over scanning positions, which indicates that the fluctuation of Te concentration along the scan is as large as 20%. Areas with Te content fluctuation around or smaller than 20% are also common. We also measured the intensity profile on the ADF image along the scanning path in Fig. 3(a) by first averaging intensity within 1 nm width along the line, followed by averaging intensity on every 1.2nm distance interval, giving the solid circle data in Fig. 3(c). This intensity profile reasonably matches the Te content fluctuation shown as in the histogram. Similar fluctuation of Te concentration around 15-20% was found in the other samples.

The chemical inhomogeneity of Te/Se content observed in STEM-EELS experiments suggests phase separation in single crystals of $Fe_{1.07}Te_{0.75}Se_{0.25}$, $Fe_{1.08}Te_{0.55}Se_{0.45}$, $FeTe_{0.7}Se_{0.3}$ and $FeTe_{0.55}Se_{0.45}$. The configuration of interconnected, fractal phases is often taken as characteristic of spinodal decomposition controlled by diffusing process with a negative diffusion coefficient, which occurs when a system makes a transition from a one-phase region at higher temperature into a spinodal region at lower temperature in the miscibility gap of the phase diagram. The average domain size $R$ was measured quantitatively by calculating $A/L$, where $A$ is the total area of the ADF image and $L$ is the total interface length[12], giving $R$ of 3.98 nm, 3.64 nm, 3.15 nm and 3.54 nm for the sample $Fe_{1.07}Te_{0.75}Se_{0.25}$, $Fe_{1.08}Te_{0.55}Se_{0.45}$, $FeTe_{0.7}Se_{0.3}$ and $FeTe_{0.55}Se_{0.45}$, respectively. However, similar interconnectivity of phases could also be achieved from a superposition, on projection, of many particle-like nanodomains, which is formed via nucleation and growth process if the system lies outside the spinodal region in the miscibility gap[13]. It is difficult to determine, by projected morphology alone, whether the origin of this phase decomposition is spinodal decomposition or nucleation and growth[13]. Nanodomains with different composition may have different transition temperature $T_c$, which has certain contribution to relatively broad superconducting transition width, $\Delta T_c$, that was recorded normally greater than 1 K and could be as large as 4 K even for high quality single crystals[3,5]. The phase separation also implies fluctuation in other physical properties, like magnetic ordering[4] which may vary among separated phases.

In summary, we investigated $Fe_{1+y}Te_xSe_{1-x}$ single crystals by using STEM and EELS techniques. Phase separation, with average domain sizes ranging from 3.1 nm to 4 nm, was found for both superconducting and non-superconducting samples, which provides microstructural evidence for broad transition temperature and implies fluctuation of related physical properties like magnetic ordering.

This material is based upon work supported as part of the Center for Emergent Superconductivity, an Energy Frontier Research Center funded by the U.S. Department of Energy, Office of Science, Office of Basic Energy Sciences. Work at the University of Illinois is supported by DOE: DE-AC0298CH1088 (H.H.) and DOE: DE-FG02-07ER46453(W.K.P.). Work at Brookhaven is supported by DOE: DEAC02-98CH10886. J.S.W and Z.J.X are supported by the Center for Emergent Superconductivity of EFRC of DOE.

Figure Captions

Fig. 1 An ADF image of FeTe$_{0.7}$Se$_{0.3}$ at atomic resolution. Te/Se atomic columns in blue are separated with Fe atomic columns in red.

Fig. 2 (a) The ADF image at a medium magnification along c axis of sample FeTe$_{0.7}$Se$_{0.3}$. (b) The magnified image from the area indicated by a square mark in (a), showing coherent interfaces at domain boundaries.

Fig. 3 Quantification of Te content by using EELS. (a) An ADF image of FeTe$_{0.7}$Se$_{0.3}$ with a line demonstrating the path along which EELS spectra were recorded. (b) The black solid curve shows a typical spectrum. The red one shows the background subtracted spectrum. (c) The histogram indicates the Te content long the scan path and the red dotted data shows the corresponding intensity profile in the ADF image.

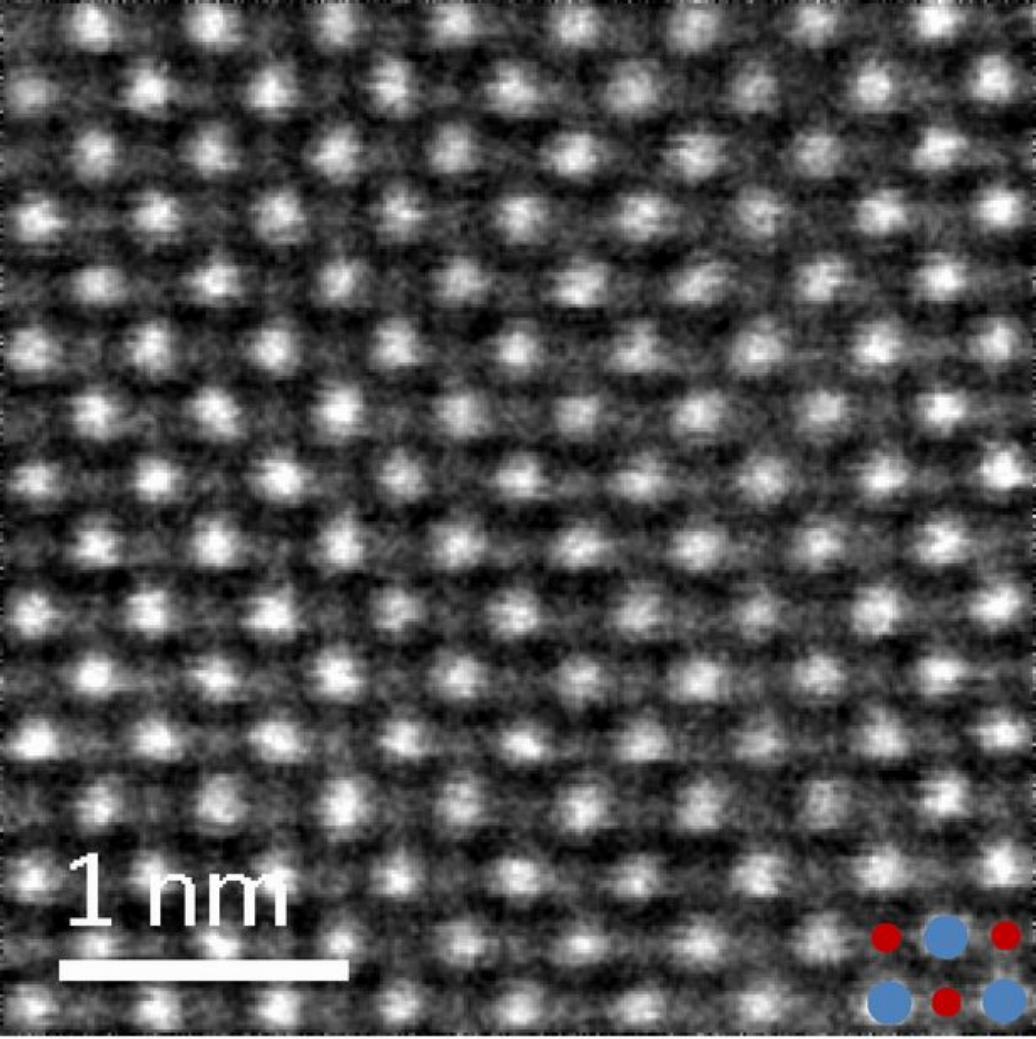

1 nm

● Te/Se (Z=52/34)   ● Fe (Z=26)

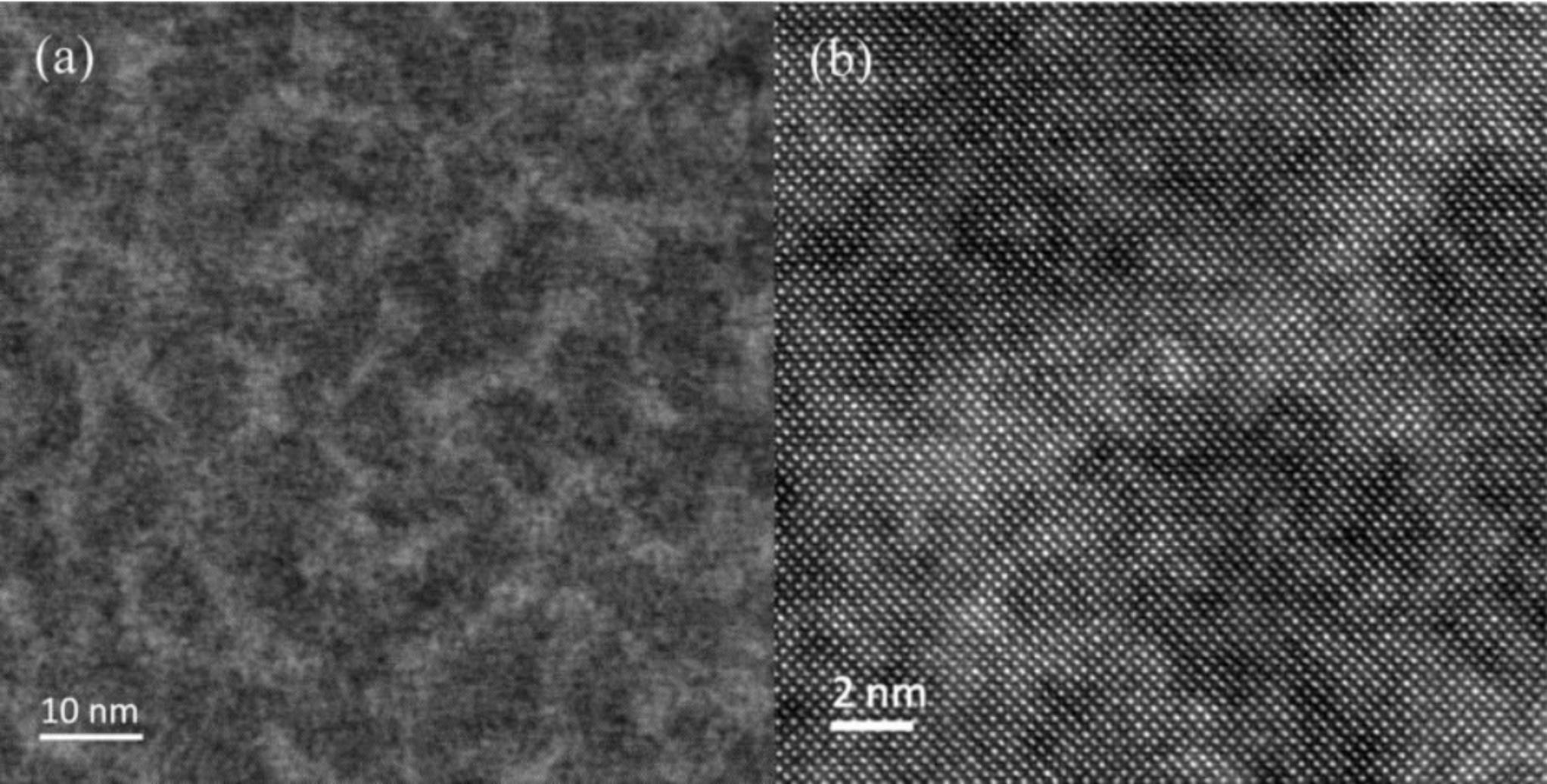

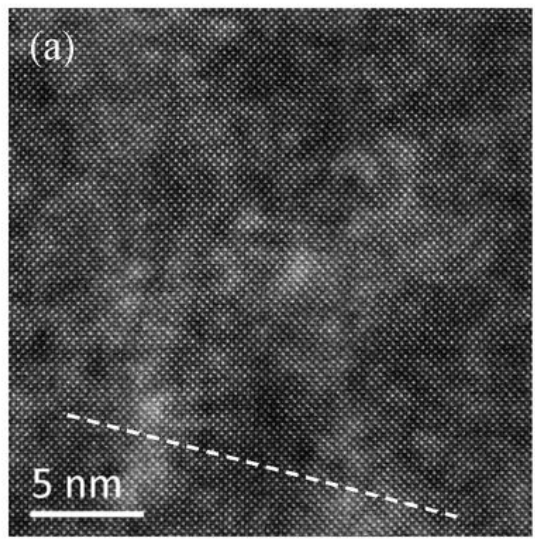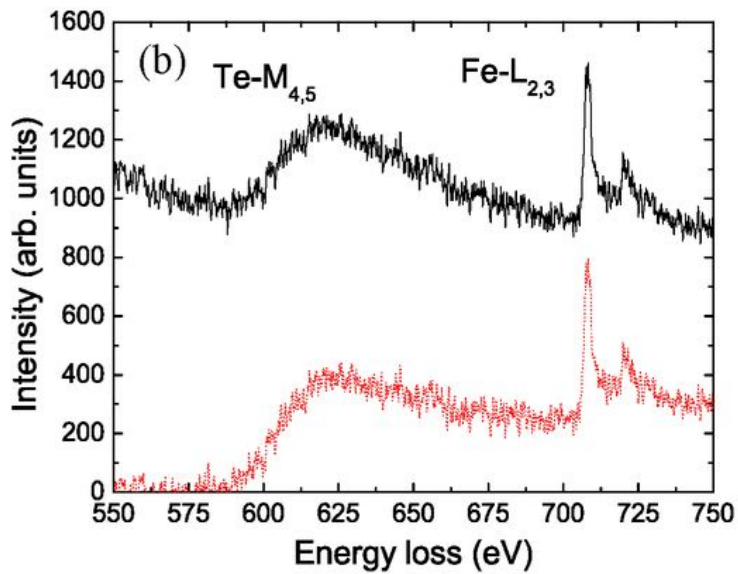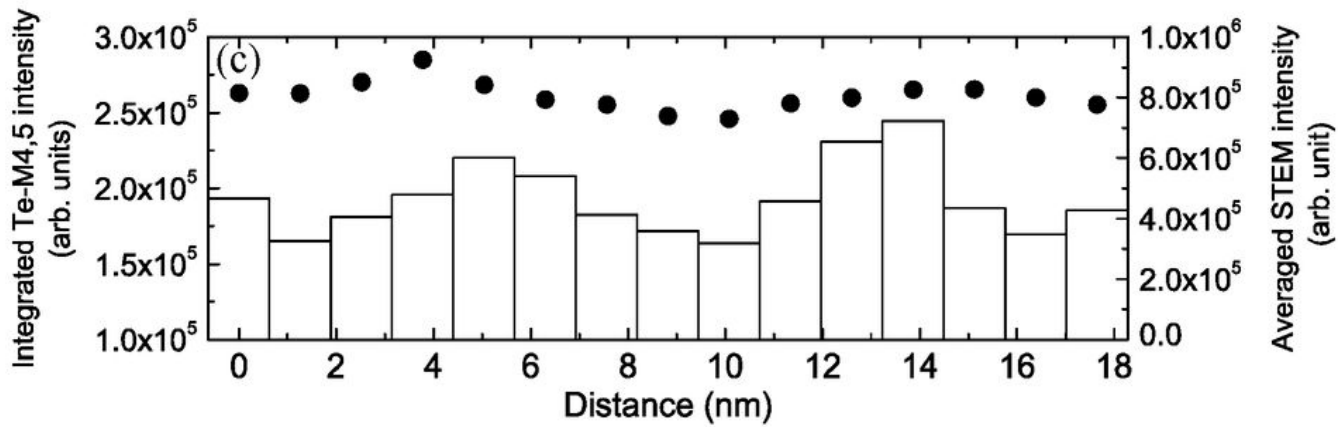